\documentstyle[12pt,aaspp4]{article}

\begin{document}

\title{Flares from the Tidal Disruption of Stars by Massive Black Holes}
\author{Andrew Ulmer\altaffilmark{1}}
\medskip
\affil{Princeton University Observatory, Peyton Hall, Princeton, NJ 08544}

\altaffiltext{1}{andrew@astro.princeton.edu}
 
\centerline{submitted to {\it ApJ}, June 24 1997}

\begin{abstract}
Tidal disruption flares are differentiated into two
classes -- those which are sub-Eddington and those which radiate
near the Eddington limit.
Flares from black holes above $\sim 2 \times 10^7 M_\odot$
will generally not radiate above the Eddington limit.
For a Schwarzschild black hole, the
maximum bolometric luminosity of a tidal disruption
is $\sim L_{\rm Edd}(5 \times 10^7 M_\odot$), substantially below the
Eddington luminosities of the most massive disrupting black holes
($\sim2\times 10^8 M_\odot$).
Bolometric corrections to the spectra of the brightest flares are found
to be large $\sim 7.5$~mag. Nevertheless, the brightest flares are likely
to have absolute magnitudes in excess of -19 in V and -21 in U
(in the absence of reddening).
Because the spectra are so blue, K-corrections may actually brighten the
flares in optical bands. If such flares are as frequent as believed,
they may soon be detected  in low or high redshift supernovae searches.
The He~II ionizing radiation produced in the flares may dominate that
which is produced by all other sources in
the centers of quiescent galaxies, creating 
a steady state, highly ionized, fossil nebula with an extent of
$\sim 1$~kpc which may be observable in recombination lines.

\end{abstract}

\keywords{accretion, accretion disks -- black hole physics -- Galaxy: center -- galaxies: nuclei -- quasars: general }

\section{Introduction}

Many lines of indirect evidence suggest that
massive black holes reside in the centers of a substantial fraction of
galaxies (e.g. \cite{res97}; \cite{kor95}; and references therein).
For example,
remarkable stellar-dynamical studies of M32 (\cite{van97})
and maser studies of NGC~4258 (\cite{wat94}; \cite{miy95})
show extreme central mass concentrations.
Such  concentrations are difficult to understand without
a black hole. The most popular alternative explanation of extremely dense
clusters, runs into problems at such high densities
(e.g. \cite{goo89}), although dynamical ``mixtures'' of a black hole with a
small, dense cluster cannot be excluded (e.g. \cite{van97}).
A nagging theoretical problem with black holes in quiescent galactic centers
was that the black hole luminosities were too low for the expected levels of
accretion; however the low luminosities can now be
understood in context of low-efficiency, advection dominated models
(e.g. \cite{nar95b}).

If black holes do, in fact, reside in the centers of galaxies, one
clinching sign would be the tidal disruption 
of stars by the black hole (e.g. \cite{res88}).
Importantly, the detection of a single tidal disruption event could even
allow for a crude determination of the black hole mass
(e.g. Loeb and Ulmer 1997).
Such estimates would yield the black hole mass apart from any contribution
from a dense stellar cluster.

Tidal disruption events are
rare ($\sim 10^{-4}$ per year per $L_\star$ galaxy), bright, short
(months to years), and seemingly unavoidable consequences
of $10^6-10^8 M_\odot$ black holes in galactic centers.  The
general picture for tidal disruption of a star has been developed over
the last 20 years (e.g. Hills 1975; Young, Shields, \& Wheeler 1977;
Rees 1988).  The picture has evolved from one in which the disruption
rate was so high as to fuel quasars to one with a 
more modest rate in which disruptions are rare transients which may
slowly grow a lower mass, $\sim 10^7 M_\odot$, black hole
(Goodman \& Lee 1989).

If a star is scattered onto a near radial orbit (inside its loss
cone) and at pericenter passes within the tidal radius, $R_{\rm t}$, of the
black hole, then the star will be disrupted as has been illustrated
by a variety of numerical simulations (e.g. \cite{nol82},
\cite{eva89}, \cite{lag93}, \cite{die97}).
For the Sun,
\begin{equation}
R_{\rm t} = R_\odot (M_{\rm bh}/M_\odot)^{1/3} \sim 25 R_s M_6^{-2/3},
\end{equation}
where $R_{\rm S}$
is the Schwarzschild radius, and $M_6$ is the black hole mass in units of
$10^6 M_\odot$.

As the star is ripped apart by the tidal forces of the black hole,
the debris is thrown onto high eccentricity orbits
with a large range of energy/period, where
\begin{equation}
\label{deltae}
\Delta E \sim {G M_{\rm bh} R_\star \over R_{\rm p}^2}
\end{equation}
and $R_{\rm p}$ is the pericenter of the star's orbit
(\cite{lac82}).
The distribution of mass as a function of energy is nearly constant as is
borne out by simulations (e.g. \cite{eva89}, \cite{lag93}).

Half of the debris is unbound with velocities around
$5 \times 10^3$~km/s.
This unbound debris
may possibly produce some long-term observable effects in the
surrounding medium. Nucleosynthesis due to the extreme compression of a
star may occur in a small fraction of tidal disruptions
(\cite{lum89}; \cite{lum90}) and lead to observable enrichment
within the Galactic center.
In AGN and Seyferts, a highly-enhanced rate of tidal disruptions may produce
enough unbound debris to form broad line and narrow line quasar-like
regions (\cite{roo92}).
Khokhlov and Melia (1996) show that
interactions between the high-velocity unbound debris 
and the ISM which may result in a supernovae-like remnant
with properties similar to those of Sgr~A~East.
In extremely low-angular-momentum encounters, the material could deposit over
$10^{52}$~ergs into the ISM, creating a very large supernova remnant.
More typically, the energy deposited would be $\sim 10^{50}$~ergs.

The bound half
may create a bright flare as it accretes onto the black hole.
The most bound material first returns to pericenter after a time,
\begin{equation}
\label{tmin}
t_{\rm min} = {2 \pi R_{\rm p}^{3} \over (GM_{\rm bh})^{1/2}
(2 R_\star)^{3/2} } \approx
0.11 \left( {R_{\rm p} \over R_{\rm t} } \right)^3
\left( {R_\star \over R_\odot} \right)^{3/2}
\left( {M_\star \over M_\odot} \right)^{-1}
M_6^{1/2}  {\rm years}.
\end{equation}

The bound material returns to pericenter at a rate (\cite{res88},
\cite{phi89}) given by
\begin{equation}
\label{mdot}
\dot{M} \sim {1 \over 3} {M_\star \over t_{\rm min}}
{\left( t \over t_{\rm min} \right)}^{-5/3}.
\end{equation}
The peak return rate according to simulations by Evans~\&~Kochanek (1989) is
\begin{equation}
\label{mdotek}
{\rm Peak~Rate} \sim 1.4
\left( {R_{\rm p} \over R_{\rm t} } \right)^{-3}
\left( {R_\star \over R_\odot} \right)^{-3/2}
\left( {M_\star \over M_\odot} \right)^{2}
M_6^{-1/2}  M_\odot~{\rm year}^{-1}.
\end{equation}
which occurs at $t \sim 1.5 t_{\rm min}$.

The scalings given above are for $R_{\rm p} \gg R_{\rm S}$. When
$R_{\rm p} \sim R_{\rm S}$, the tidal interactions are
stronger (e.g. \cite{fro94}). These effects of the Schwarzschild
metric can be approximated in
the pseudo-Newtonian potential of Paczy\'nski and Wiita (1980),
$\Psi(R) = - GM/(R - R_{\rm S})$.
For instance, close to a black hole, the tidal radius,
$R_{\rm t,S} = R_\odot (M_{\rm bh}/M_\odot)^{1/3} + R_{\rm S}$,
differs from that of the Newtonian potential.
The deepest, non-captured orbit has $R_{\rm p} = 2 R_{\rm S}$,
so as is well-known, the maximum mass black hole that can disrupt
a solar-type star is $\sim 1 \times 10^8 M_\odot$. As
discussed in \S~IV, encounters more
distant than the tidal radius may still strip significant matter from the
star allowing disruptions for black holes up to $\sim 2 \times 10^8 M_\odot$
in the Schwarzschild case.
Also altered is the range of energy of the debris (Eq.~\ref{deltae}),
\begin{equation}
\label{deltaes}
\Delta E_{\rm S} \sim {G M_{\rm bh} R_\star \over (R_{\rm p}-R_{\rm S})^2}.
\end{equation}
The increase in binding requires that Eqs.~\ref{tmin}-\ref{mdotek} be
modified (for the pseudo-Newtonian potential, $R_{\rm p}$ is replaced
by $(R_{\rm p}-R_{\rm S})$), and has the effect of reducing the
minimum return time and of increasing the peak debris return rate.

Another relativistic phenomenon, orbital precession, may lead to the
crossing of debris orbits and may form strong shocks (\cite{res88}).
The explicit evolution of the debris is complex and is dependent on the
black hole's properties as well as those of the star and the stellar orbit.
As discussed by Kochanek (1994), at intersections between the debris streams,
energy and angular momentum may be transfered. The computational
complexities of the problem prohibit a direct numerical solution, but
as an approximation, a large fraction of the debris can be said to
circularize in a time,
\begin{equation}
\label{tcir}
t_{\rm cir} = n_{\rm orb} t_{\rm min},
\end{equation}
where $n_{\rm orb}$ is a number greater than 1, and probably between 2 and 10.

After, and perhaps while, the debris circularizes, it may accrete onto the
black hole. The dynamical time at the tidal radius is short relative to
the return time, so it is likely that the disk accretes quickly, resulting in
a luminous flare. Cannizzo, Goodman, \& Lee (1990) investigated
the accretion of a tidally disrupted star in the thin disk case.

Ulmer (1997) has investigated accretion in a thick disk case.
The emergent spectra may have a strong dependence on 
the structure and orientation of the disk, as well as on
the possible existence of winds.
A starting point for models of the emergent radiation is the
temperature associated with an Eddington luminosity emanating from
the tidal radius
\begin{equation}
\label{teff1}
T_{\rm eff} \approx \left( { L_{\rm E} \over 4 \pi R^2_{\rm t} \sigma}
\right)^{1/4}
\approx 2.5 \times 10^{5} M_6^{1/12} \left( {R_\star \over R_\odot}
\right)^{-1/2} \left( {M_\star \over M_\odot}
\right)^{-1/6} K
\end{equation}
or from a small factor times the Schwarzschild radius:
\begin{equation}
\label{teff2}
T_{\rm eff} \approx \left( { L_{\rm E} \over 4 \pi (5 R_{\rm s})^2 \sigma}
\right)^{1/4}
\approx 5 \times 10^{5} M_6^{-1/4} K.
\end{equation}
In either case, the temperature scaling with black hole mass is slight.
Therefore, in the simplest models, all disruption flares would have comparable
characteristic temperatures.

A tidal disruption event has yet to be definitively detected, but there
are a number of suggestive observations.
First, a UV flare from the center of an elliptical
was observed on the timescale of a year
by Renzini~et~al. (1995), but this flare is $10^{-4}$ as luminous as
might be expected. Furthermore, the spectrum flattens in the near UV
($\sim 2000 \AA$) which would not be expected except perhaps
in the case of a very strong wind (see \S~V). The possibility remains,
as they suggest, that they have observed the disruption of only the
outer layer of a star.
Second, Peterson and Ferland (1986) observed the
optical brightening of an optically variable Seyfert along with the
appearance of a strong, abnormally broad He~II recombination line.
The event is intriguing, but without a light curve or other definitive
evidence, the event is not conclusive.
As pointed out by Rees (1997), the most convincing disruption would be one
which comes from a quiescent galaxy.
Third, observations of variable double-peaked H$_\alpha$ lines in spectrum of
Seyfert NGC~1097 may be modeled as emission lines from the appearance of an
elliptical disk (possibly formed through tidal disruption)
close to a black hole, but other explanations exist
(Eracleous~et~al. 1995; Storchi-Bergmann~et~al. 1995).

The outline of the paper is as follows. In \S~II, a
crude, but useful phenomenology of accretion disks is discussed.
In \S~III, time scales in the tidal disruption event are presented.
In \S~IV, these time scales are used to delineate regions of
Eddington-level tidal disruption flares from the generally, less luminous
sub-Eddington flares.
In \S~V, characteristics of the emergent spectra from these flares are
investigated. Conclusions are given in \S~VI.

\section{Accretion Disk Phenomenology \label{lumsec}}

In this section, the known relationships between luminosity
and mass accretion rate are examined for a variety of accretion regimes
which are important in determining the luminosity of tidal
disruption flares. The mass accretion rate is given in
terms of the Eddington accretion rate, $\dot{M}_{\rm {E}} \equiv
{L_{\rm Edd} /\epsilon c^2}$, where $\epsilon$ is the efficiency
of the accretion process and is taken to be 10\%.

\begin{figure}
\plotone{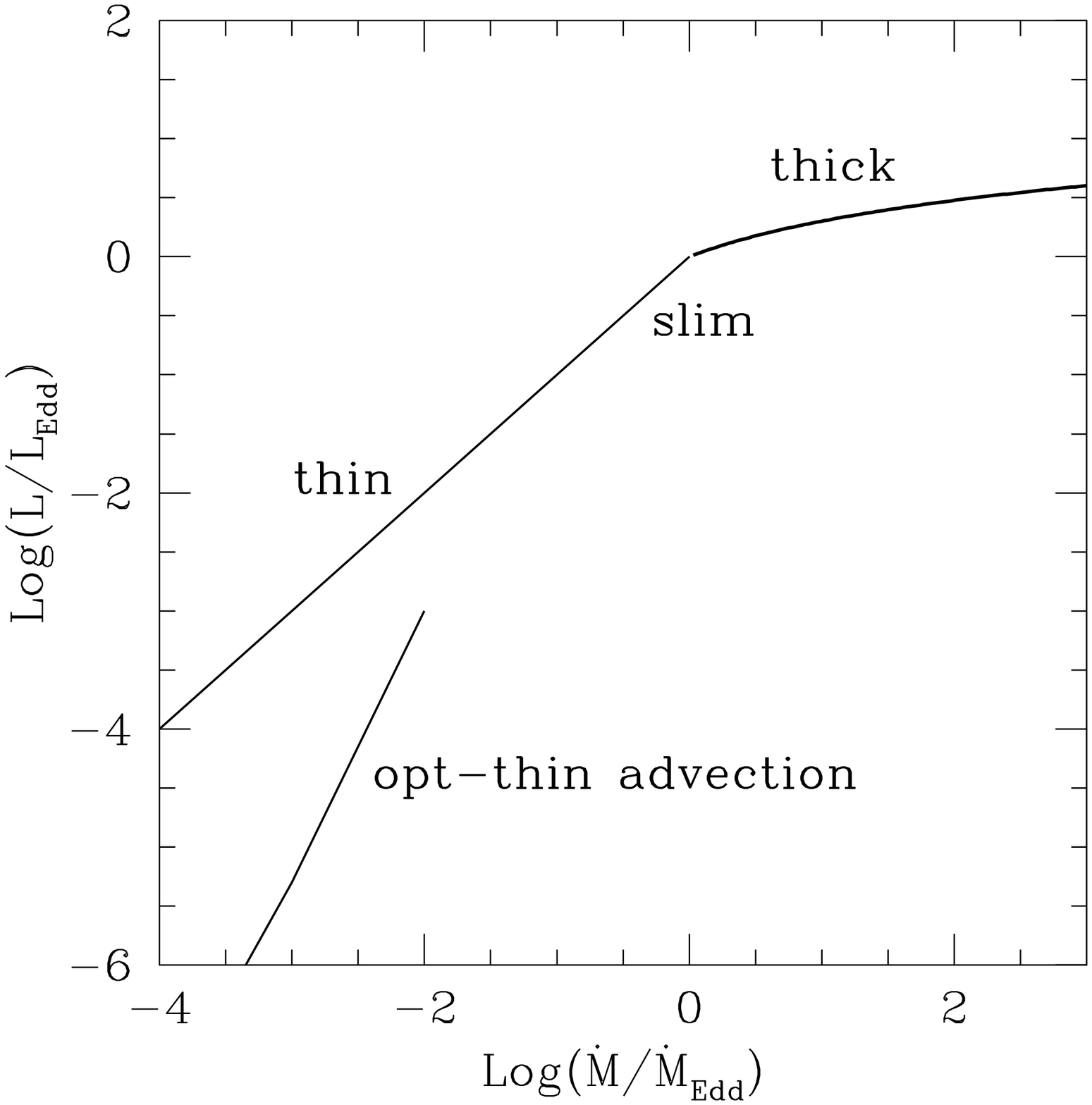}
\caption{\label{accfig}
Schematic of luminosity versus mass accretion rate
(where $\dot{M}_{\rm Edd} \equiv {L_{\rm Edd} /\epsilon c^2}$),
where $\epsilon = 0.1$ is the efficiency of the accretion process.
The brightest accretion disks have high mass accretion rates and are thick.
Thick disks are not much brighter than the Eddington luminosity,
because at very high accretion rates
pressure gradients push the inner edge of the disk closer to the marginally
bound orbit, and energy is advected (because of geometry)
(e.g. Paczynski and Wiita, 1980). Thin disks, (e.g. Shakura \& Sunyaev 1973)
and optically thin advection dominated models (e.g. Narayan \& Yi 1995) are
less luminous.}
\end{figure}

Figure~\ref{accfig} shows a schematic of luminosity versus mass accretion rate.
The brightest known accretion disks have high mass accretion rates and are
optically and geometrically thick, but thick disks are never
much brighter than the Eddington luminosity,
because at very high accretion rates,
pressure gradients push the inner edge of the disk closer to the marginally
bound orbit, so that less binding energy is released during accretion and
more is carried into the black hole (e.g. \cite{pac80b}).
The luminosity increases logarithmically with mass accretion rate for thick
disks (\cite{pac80a}).
Thin disks are less luminous, with luminosity scaling linearly
with mass accretion rate (e.g. \cite{sha73}).
Slim accretion disks (\cite{abr88}; \cite{szu95}),
may form a transition between the thin and
thick accretion regime.
At very low accretion rates, optically thin advection dominated models
may exist and are even less luminous (Narayan \& Yi 1995).
They exist at a maximum accretion rate of $\sim \alpha^2\dot{M}_{\rm E}$,
and have $L\sim \dot{M}^2$,
so that the luminosity falls off faster with decreasing
accretion rate.
A possibility exists that at accretion rates very much higher than
the Eddington rate (and probably beyond those produced by stellar
disruptions), the accretion may become more spherically symmetric and
highly optically thick, so the radiative diffusion time could exceed the
the infall time. In this case, much of the binding energy
could be carried into the black hole
(e.g. \cite{shv71}; \cite{ips77}; \cite{fla84})
and the luminosity could actually fall below the Eddington limit.
An additional
complication is the possibility of beaming in the thick disk regime,
but for simplicity and because of the absence of a preferred beaming
model, beaming is set aside for the time being.

The simplified view of accretion disks is that the luminosity increases 
with mass accretion rate until the Eddington mass-accretion rate,
at which point the luminosity reaches the Eddington limit and
remains at that level for all higher mass accretion rates.
Accretion events with sub-Eddington luminosities occur in thin disks,
and accretion proceeds in thick disks for all super-Eddington accretion
rates. For sub-Eddington
accretion rates, the luminosity falls off at least as fast as linearly with
mass accretion rate.

\section{Derivation of Timescales}

Although many of the results derived in this section are already known, I
briefly rederive them here for completeness and to scale all
relevant parameters. The timescales considered are (1) the radiation time,
the time to radiate, at the Eddington limit, the energy of the bound debris;
(2) the circularization time,
the time for a large fraction of the debris to circularize; and
(3) the accretion time, the time to
accrete a large fraction of the disk.
Throughout the paper, $m$ is defined as the mass of the
star in units of the solar mass, $r$ as the radius of the star in units of
the solar radius, $M_6$ as the black hole mass in units of $10^6 M_\odot$,
$R_{\rm S}$ as the Schwarzschild radius, $2 G M_{\rm bh} / c^2$,
and $R_{\rm p}$ as
the pericenter of the orbit in Schwarzschild coordinates.

The main timescales of the problem can now be written in terms of the
system parameters.

The time to accrete all of the bound debris (which is half of the stellar mass)
if the black hole radiates at the Eddington luminosity is
\begin{equation}
\label{trad}
t_{\rm rad} = {0.5 m M_\odot c^2 \epsilon \over L_{\rm E} }
\approx 21 m M_6^{-1} \epsilon_{0.1} {\rm years}
\end{equation}
for a hydrogen mass fraction, $X=0.74$, and efficiency, $\epsilon =0.1$.

As discussed in \S I, the debris are given a range of periods, and
the circularization time is closely related to the the orbital time
of the most tightly bound debris, $t_{\rm min}$,
(see Eq. \ref{tmin} and \ref{tcir}):
\begin{equation}
\label{tcir2}
t_{\rm cir} \approx n_{\rm orb} t_{\rm min} \approx
 0.11 n_{\rm orb} \left( {R_{\rm p} \over R_{\rm t} } \right)^3
r^{3/2}
m^{-1}
M_6^{1/2}  {\rm years},
\end{equation}
where $n$ is the small number of orbits necessary for circularization to
occur.

The time in which a large fraction of the matter in a disk can accrete
can be written
\begin{eqnarray}
\label{tacc}
t_{\rm acc} & \approx & {2 R_{\rm p} \over v_{\rm r}} \\
& \approx & {t_{\rm Kep}(2 R_{\rm p}) \over \alpha \pi h^2} \\
&\approx & {3 \times 10^{-4} \over \alpha h^2}
\left({R_{\rm p} \over R_{\rm t} }\right)^{3/2} r^{3/2}
m^{-1/2}  {\rm years},
\end{eqnarray}
where $h$ is the ratio of disk height to radius and is approximately one for
thick disks, and $t_{\rm Kep}$ is the rotation period in a Keplerian
potential. In \S~\ref{lumsec}
the geometrical thickness of the disk is discussed,
but unless noted otherwise, it is the thick disk case that is of interest.

\section{Comparison of Timescales}

By considering the main timescales of tidal disruption events,
the gross characteristics of the flares are investigated.
Of particular interest
is the luminosity and whether
the flare is super-Eddington or sub-Eddington. The argument that will
be made stems from
the system's time scales:
$t_{\rm rad}$, the time to radiate, at the Eddington limit, the energy of
the bound debris;
$t_{\rm cir}$, the time for a large fraction of the debris to circularize; and
$t_{\rm acc}$, the time to
accrete a large fraction of the disk.

Consider the three possibilities: (1) the accretion rate is so high
that the material is drained from the disk nearly as fast as it circularizes,
(2) the accretion rate is slower than circularization but faster than the
radiation time so the material forms a reservoir which accretes in a
thick disk,
(3) the material accretes at a rate slower than it circularizes, and
on a time scale longer than the radiation time, so the disk is not
thick and the debris accretes in the less luminous form of a thin disk.

Let us first consider the these scenarios for a
solar-type star with pericenter at the tidal radius which is disrupted by
a $10^6 M_\odot$ black hole. The parameter, $n_{\rm orb}$, related to
the efficiency of circularization, is taken as 2.
The only unspecified parameter is $\alpha$.
For a $10^6 M_\odot$ black hole:
(1) if $\alpha \gtrsim 1.5 \times 10^{-3}$, then
$t_{\rm acc} \lesssim  t_{\rm cir}$ and a thick disk will form and drain as
the material circularizes;
(2) if $1.5 \times 10^{-5} \lesssim \alpha \lesssim 1.5\times 10^{-3}$,
then $t_{\rm cir} \lesssim t_{\rm acc}  \lesssim t_{\rm rad}$ and a 
a reservoir of debris will form and accrete in a thick disk; and
(3)if $\alpha \lesssim 1.5 \times 10^{-5}$, then $t_{\rm acc} \gtrsim
t_{\rm rad}$ and a reservoir of debris will form and accrete in a thin disk.
Therefore, unless $\alpha$ is extremely small, much smaller than
($\sim 10^{-1} - 10^{-3}$) as is
estimated from, for example, DN and FU outbursts
(\cite{can92}; \cite{lin96}; and references therein)
and from numerical experiments of the
Balbus-Hawley mechanism (\cite{haw95}; \cite{sto96}),
thick disks are the norm.
Therefore, it appears likely that for a $10^6 M_\odot$ black hole, the
tidal disruption of a star produces a thick disk which will radiate near
the Eddington limit (although it is still uncertain in what energy band the
energy will be released).

\begin{figure}
\plotone{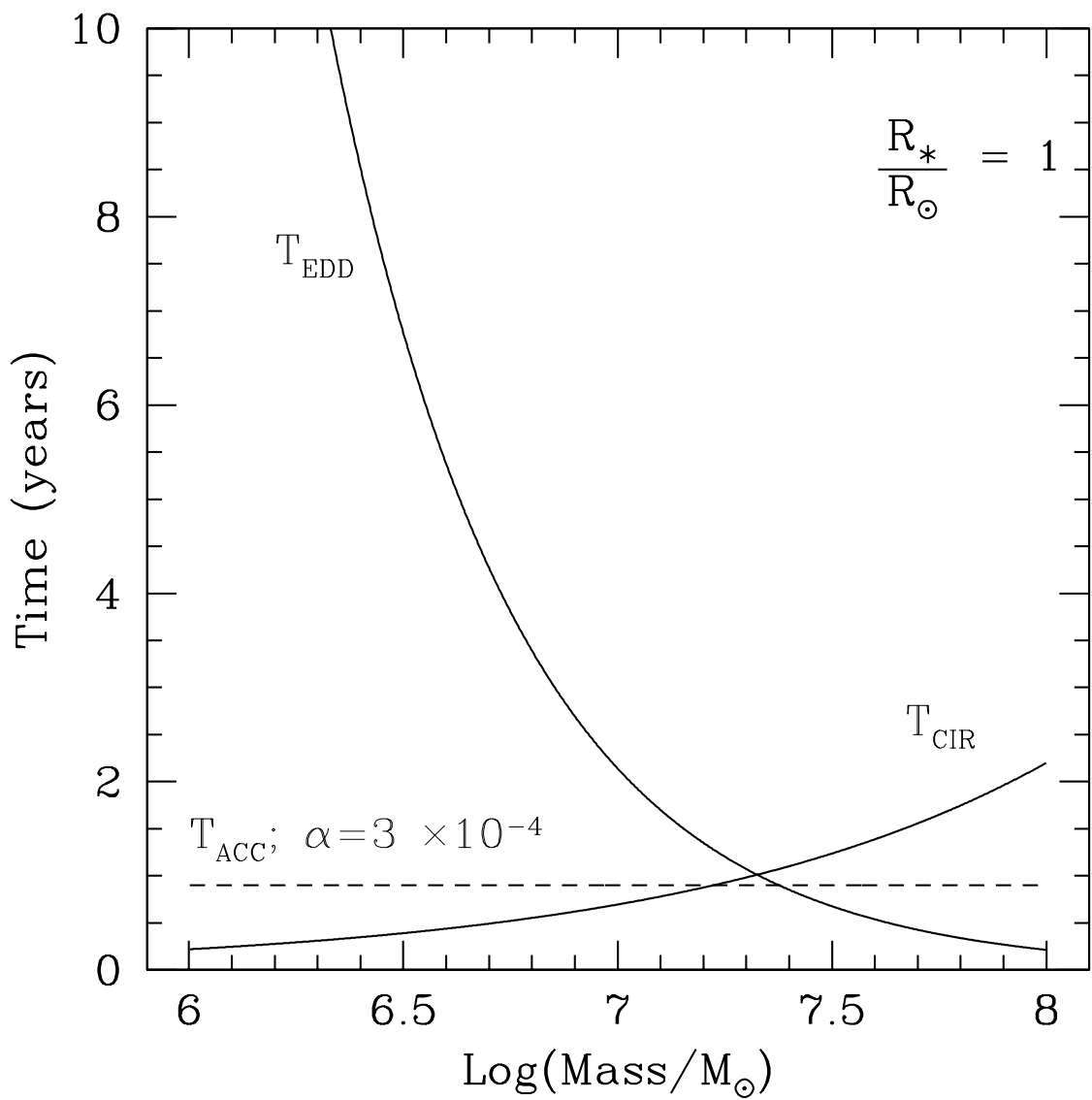}
\caption{\label{tscalefig}
Timescales as a function of black hole mass
a solar-type star with a pericenter equal to the tidal radius.
For a $10^6 M_\odot$,
black hole, thick disks can form and bright flares can
occur if $t_{\rm acc}$ is
less than $t_{\rm cir}$ which requires
only that $\alpha > 1.5 \times 10^{-5}$.
Above $\sim 10^7 M_\odot$, the Eddington time becomes shorter than the
circularization time, so the disk cannot be thick, and the event will be less
luminous.}
\end{figure}

In figure~\ref{tscalefig}, these timescales are shown
as a function of black hole
mass, again for the case of
a solar-type star with a pericenter equal to the tidal radius.
Because of the different scalings with black hole mass, the circularization
time becomes longer than the radiation time at $M_{\rm bh} > 2
\times 10^7 \epsilon_{0.1}^{2/3} M_\odot$.
This finding is independent of $\alpha$ as long as $\alpha$ is larger
than $\sim 1.5\times 10^{-4}$.
When the peak accretion rate falls below the Eddington
rate above this critical mass,
the accretion should occur in a slim disk (\cite{abr88})
or a thin disk rather than a thick disk.
Although little has been said yet about the spectra,
it is clear that there may be observational differences between the
two classes.

A similar conclusion may be reached by considering when the
peak return rate of the stellar debris (eqs. \ref{mdotek}) is super-Eddington.
For an accretion efficiency of $0.1$, the maximum mass
is $1.5 \times 10^7 M_\odot$.
This estimate is slightly smaller than that
obtained by considering the timescales, because the circularization time
is an underestimate,
since a fraction of the matter returns in times longer than the
circularization time.
Nevertheless, both estimates point to a
transition mass around $10^7 M_\odot$.
For more massive black holes, the
return rate is lower, and the luminosity, which for slim to thin disks
scales roughly linearly with mass accretion rate, will be lower than Eddington.

The treatment above has be carried out for a solar type star with
pericenter equal to the tidal radius. Let us first consider the
results for a less massive, main sequence star.
These stars will generally be denser with
$r \approx m^{0.8}$ (e.g. \cite{kip90}). Then, the transition mass
will be $\sim 2 \times 10^7 m^{0.8} M_\odot$.

The last parameter that will be considered is
the pericenter of the orbit which for disruptions may be small as
$2 R_{\rm S}$ in Schwarzschild coordinates. The upper radius, or
the disruption radius, $R_{\rm d}$, is the radius at which a star in a
parabolic orbit is barely disrupted by the tidal field of a black hole.
The disruption radius is of order the tidal radius, and is given by
\begin{equation}
\label{disrad}
R_{\rm d} = q R_{\rm t},
\end{equation}
where the parameter, q, is between about 1 and 3 and
hides a number of complexities
including relativistic corrections to the potential.
Throughout the paper, q is taken as 2.
Young, Shields, and
Wheeler (1977); Luminet and Carter (1986);
Khokhlov, Novikov, and Pethick (1993);
Frolov~et~al. (1994);
and Diener~et~al. (1997)  contain discussions of this
parameter with a range of analytical
approximations and numerical methods.
The demarcation of $R_{\rm d}$ is hazy as stars still may lose a fraction of their
mass at larger pericenteric orbits.

According to the theory of loss cones (\cite{fra76}; \cite{lig77})
there is a critical orbital radius outside of which 
a star's orbit diffuses in one orbit
at an angle greater than the loss-cone angle.
If stars which diffuse onto the loss cone primarily come from inside
this critical radius, there will be an excess of events with
$R_{\rm p} \sim R_{\rm d}$. If stars diffuse primarily
from large radii, where
the diffusion angle is larger than the loss cone, one can use the
``$n\sigma v$ approximation'' which Hills (1975) used to calculate
the tidal disruption rate to show that the distribution of lost stars
is constant with $\omega$ (gravitational focusing cancels
the gain in cross section at larger $\omega$).
The true distribution of lost stars will be a combination of the two extremes
mentioned, with an excess of events near $R_{\rm p} = R_{\rm d}$, and a tail
towards $R_{\rm p} = 2 R_{\rm S}$, corresponding to ``plunge-orbits''
in a Schwarzschild metric (\cite{nov73}).
The plunge orbits, in which the star is directly swallowed by the
black hole, will not produce any tidal debris or accompanying flare.
It is well-known that above $\sim 2\times 10^8 M_\odot$, the plunge
radius surpasses the disruption radius, so no tidal disruption
events can be produced.

\begin{figure}
\plotone{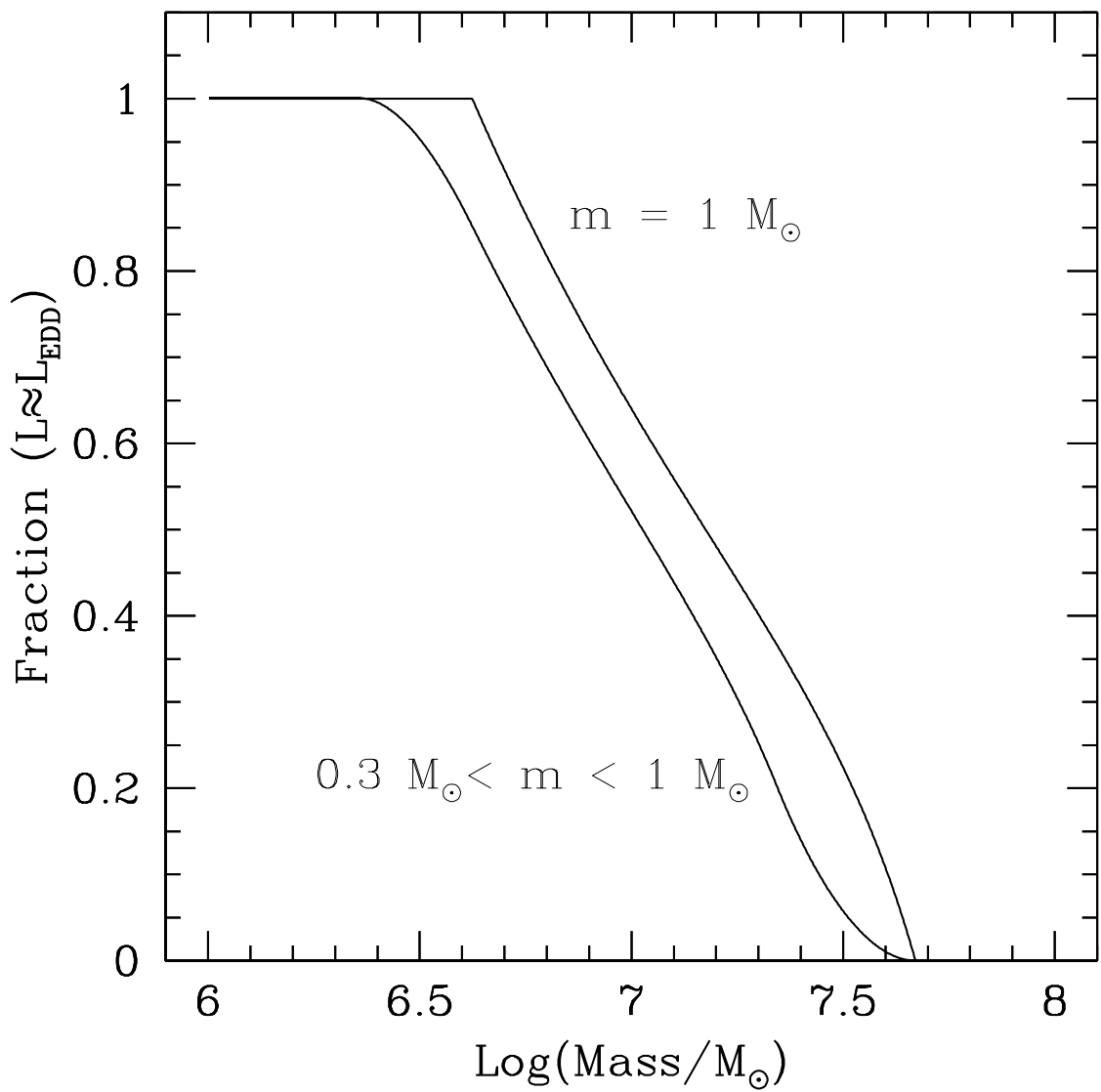}
\caption{ \label{fracfig}
The fraction of tidal disruption events with
peak accretion rates greater than the Eddington-limit are shown
(upper curve) for the case of solar-type stars with
a uniform distribution of pericenters and (lower curve) stars with
a Salpeter mass function between $0.3 M_\odot$ and $1 M_\odot$,
also with a uniform distribution of pericenters.}
\end{figure}

Tidal disruption events with smaller pericenters will have higher
peak luminosities than those with $R_{\rm p} \sim R_{\rm t}$
(see Eq. \ref{mdotek}).
Therefore, the actual transition mass discussed above is better
regarded as a gradual rather than a sharp transition.
In figure~\ref{fracfig}, the fraction of tidal disruption events with
peak accretion rates greater than the Eddington-limit is shown for
the case of a uniform distribution of $R_{\rm p}$ and stars with
a Salpeter mass function between $0.3 M_\odot$ and $1 M_\odot$.
More realistic distributions of $R_{\rm p}$, formed by integrating over
a cusp, would provide a somewhat sharper transition.

An interesting result of these scaling arguments is that if  the
luminosity is roughly limited at the Eddington luminosity, then there is
a maximum luminosity that a tidal disruption event may produce.
This luminosity occurs when the peak mass return rate is
the Eddington rate and the pericenter is the minimum ($2 R_{\rm S}$)
for a tidal disruption. The maximum luminosity corresponds to
the Eddington limit for a black hole mass
$\sim 5 \times 10^7 \epsilon_{0.1}^{1/3} M_\odot $, i.e.
$7 \times 10^{45}$~ergs~s$^{-1}$ and $2 \times 10^{12} L_\odot$.
The maximum luminosity is probably somewhat lower, because the
estimate assumes that all of the gravitational binding energy of the
returning debris is released as it returns to pericenter, whereas
the circularization process is probably slower.
Such extreme events would be rare, because of the small range of allowable
impact parameters.
The peak luminosity could be higher for tidal disruption events
which occur in the orbital plane of a Kerr black hole.

Following Rees (1988) and Evans and Kochanek (1989), the duration of
a flare may be calculated based on the time that the infall rate is
super-Eddington with the assumption that the circularization and
accretion are rapid.
The duration is
\begin{eqnarray}
\label{dur}
t_{\rm sup} & \approx & t_{\rm min} 
\left[
18
\left( {R_{\rm p} \over R_{\rm t} } \right)^{-9/5}
r^{-9/10}
m^{6/5} \epsilon_{0.1}^{3/5}
M_6^{-9/10}  - 1.5
\right] \\
& \approx & 1.9
\left( {R_{\rm p} \over R_{\rm t} } \right)^{6/5}
r^{3/5}
m^{1/5} \epsilon_{0.1}^{3/5}
M_6^{-2/5} {\rm years}.
\end{eqnarray}
If the viscosity is high enough that the matter has been
accreted roughly as fast as it circularizes, the emission should fall
off after $t_{\rm sup}$.

\section{Flare Energy Spectrum and Absolute Magnitudes}

The peak bolometric luminosity of the flares is determined
from the scaling arguments given above.
According to this scenario, the brightest flares are those in which
accretion occurs in a thick disk (or in a slim disk), so
a study of the spectrum for that case alone is the main focus.
Cannizzo, Lee, and Goodman (1990) have investigated the spectrum a flare
if the debris forms a thin disk. At the earliest and brightest times,
the spectrum was found to peak at $\sim 300 \AA$, with U and V absolute
magnitudes of -16.7 and -15.2 for a $10^7 M_\odot$ black hole.

\begin{figure}
\plotone{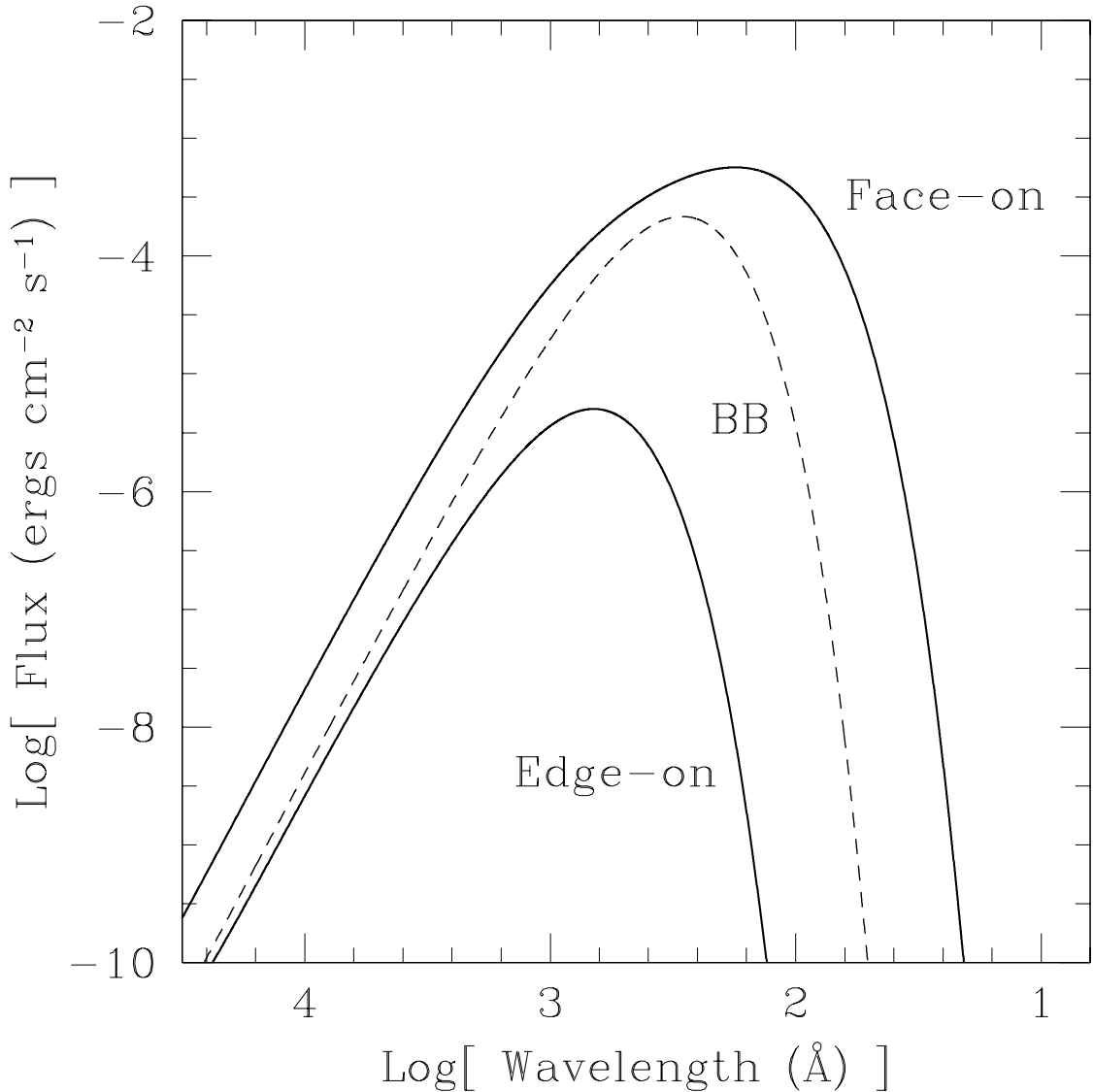}
\caption{\label{specfig}
Predicted spectra of a tidal disruption flare around a $10^7 M_\odot$
black hole. The flux is appropriate for a source at a distance of 100 Mpc.
The two solid lines show calculated spectra from an optically and
geometrically thick disk as viewed from face-on and edge-on.
The former is much brighter and hotter because a large fraction of the
energy is radiated in the funnel.  The dashed line is a Planck curve with
$T=10^5$ for comparison.}
\end{figure}

Consider a thick disk spectra to begin with.
Figure~\ref{specfig} shows the spectra calculated using a thick disk code
(Ulmer, 1997).
The spectra presented are typical thick disk spectra
for a solar mass star disrupted at $\sim 2 R_{\rm t}$,
the disruption radius (Eq. \ref{disrad}), by a $10^7 M_\odot$ black hole.
Viewing angle will have a large effect on the spectra, because when seen
edge-on, the bright, hot funnel is obscured by the disk.
Viewed face-on, the disk is very bright, radiating slightly above the
Eddington limit.
When viewed edge on, the bolometric luminosity is
less by a factor of $\sim 50$.
For the thick disks considered here, the luminosity is 2 to 3 times the
Eddington limit.
The absolute magnitudes for a disk around a $10^7 M_\odot$ black hole
radiating at 2.5 times the Eddington limit, are for face-on and edge-on
disks: U:-21, V:-19.5 and U:-19, V:-17.5, respectively.
For a similarly bright disk around a  $10^6 M_\odot$ black hole,
the absolute magnitudes are U:-17, V:-15.5 and U:-15, V:-13.5.
The bolometric corrections are $\sim -7.5$~mag in V, where
reddening, which may well be important in the center
of a galaxy, and dimming from cosmological effects (though in a flat
cosmology, very high redshift flares would brighten as $(1+z)$ due to
K-corrections) are neglected.
The disks may radiate closer to the Eddington limit than
do the disk models above which radiate at 2 to 3 times the Eddington limit.

A thick disk may drive a strong wind. If a wind were dense enough,
the effective photosphere would be moved outwards from the surface
of the disk, and the spectra would appear redder.
Some variable stars, namely luminous blue variables,
show what are believed to be wind induced color changes, resulting
in V-band brightening of 2 magnitudes (e.g. \cite{hum94}).
In super-Eddington thick disks, strong winds are to be expected on
the purely phenomenological
grounds that in stars, winds increase as a star's luminosity approaches the
Eddington limit.
Additionally, instabilities in the funnel of thick disks may drive
strong winds as well as jets (e.g. \cite{jar80}; \cite{nit82}).
Alternatively, a strong wind could be driven before the debris
completely settles into a disk. Rees (1988) observed that the
total energy required to eject the majority of the bound material
is small (see Eq.~\ref{deltae})
$\sim M_\odot c^2 R_\star R_{\rm S}/ R_{\rm p}^2$.

Limits on the strength of the wind are now examined.
The mass loss rate is limited by the energy required to lift the
material from the disk and carry it to infinity. In the context
neutron stars, Paczy\'nski \& Proszynski (1986), have considered
energy balance limits on winds.
Following Paczy\'nski \& Proszynski, the mass loss rate is
\begin{eqnarray}
\dot{m} & \approx& { (L_{\rm b}-L_{\rm ph}) 2 R_{\rm b} \over c^2 R_{\rm S}}\\
     &  \lesssim  & {L_{\rm Edd} 2 R_{\rm b} \over c^2 R_{\rm S} }
\end{eqnarray}
where $R_{\rm b}$ is the base radius from which the matter is lifted into
the wind by the radiation, and $L_{b}$ and $L_{\rm ph}$ are the luminosity at
the base radius and at the photosphere.
If the base radius is taken to be $\sim 10 R_{\rm S}$, $\dot{m} \lesssim
0.05 M_\odot  M_6 {\rm years}^{-1}$. The wind may, therefore, carry away a
sizable fraction of the disk mass for a $10^7 M_\odot$ black hole.

Given a maximal mass loss rate, the maximum photospheric radius may be
estimated as follows.
Near the photosphere, $\dot{m} = 4 \pi v \rho R_{\rm ph}^2$
and $\rho R_{\rm ph} \sim m_{\rm p}/\sigma_{\rm T}$, where
$m_{\rm p}$ is the proton mass, and $\sigma_{\rm T}$ is the Thompson
cross-section.
If $v \sim v_{\rm esc} \sim \sqrt{G M_{\rm BH} / R_{\rm ph}}$,
then
\begin{eqnarray}
R_{\rm phot} & \lesssim & {\dot{m}^2 \sigma_{\rm T}^2 \over 32
\pi^2 GM_{\rm BH} m_{\rm p}^2} \\
& \lesssim & 120 R_{\rm S} \left({v \over v_{\rm esc}}\right)^{-2}.
\end{eqnarray}
The minimum effective temperature is expected to be $\sim1.1 \times
10^5$~K, which is still very hot.
The velocity of the outgoing wind could be lower than the escape velocity,
and if the luminosity is very close to the Eddington limit at the
photosphere, the escape velocity would be reduced by
$1-L/L_{\rm Edd}$ from the expression given above, creating a larger
photosphere.
Including a photosphere with $T\sim10^5~K$
which radiates at the Eddington-limit in
the thick disk spectra calculated
above, damps
the high energy spectrum but
has little effect on the U and V band luminosities.
The optical bands are weakly affected even though a larger fraction of
the light goes into optical, primarily because the total luminosity
of the source is diminished as energy is carried away by the wind
and because a significant fraction of the optical light comes from 
part of the disk outside the photosphere at a few hundred $R_{\rm S}$.
The optical luminosities of a thick disk and of
a thick disk with a wind are nearly the same. The main effects of a strong
wind are the reduction of the very high energy spectrum of the flare and a
possible shortening of the flare by loss of the stellar debris to the wind.
Such strong winds would greatly reduce the expected number of
soft X-ray tidal disruption flares (Sembay \& West 1993) accessible,
for instance, to ROSAT.

A third possibility, and the most optically luminous one, is that
an extended, static envelope could be formed around the disk
(\cite{loe97}).
Such an envelope might form either if
the initial collisions of the debris streams or subsequent evolution
of the disk stream expels much of the matter to 
large radii, and strong radiation pressure
disperses this marginally bound gas into a quasi-spherical configuration.
In this scenario the object is spherically symmetric with Eddington
luminosity and effective temperature
\begin{equation}
T_{\rm eff}\approx 2.3\times 10^4~{\rm K} \left({M_{\rm bh}\over
10^7 M_{\rm env}}\right)^{1/4}.
\label{envteff}
\end{equation}
where $M_{\rm env}$ is the mass in the envelope. In this case, the bolometric
corrections would be much less ($\sim 2.5$),
and the objects would be as bright as Seyferts.

In addition to observing the very blue continuum of the thick disk, it
may be possible to observe a flare in emission-lines.
The disk may radiate in emission lines, but to what extent is
uncertain because the emission would depend on 
the temperature and density structure both of the disk photosphere and of
any wind. As discussed in Roos (1992) and Kochanek (1994), prior
to circularization, the bound material generally has
densities higher than $10^{12} {\rm  atoms~cm}^{-3}$,
so most of the energy the material
radiates (or absorbs and re-radiates) will be
in the continuum and not in emission lines. Moreover, such
debris has a relatively small covering fraction.
Recombination lines, e.g. H$_\alpha$, might be radiated
from the recombination of the debris or from the
ionization and subsequent recombination of a neutral
surrounding environment.
The possibility that
the disk may radiate in recombination lines was suggested by
Eracleous~et~al.~(1995) and Storchi-Bergmann~et~al. (1995),
to explain observations of variable H$_\alpha$ lines
in spectrum of the Seyfert NGC~1097.
However, the disk is probably
extremely optically thick even at thousands of $R_s$, with
typical surface density $0.5 M_\odot /(\pi (1000 R_s)^2)$ of
$3000 M_6^{-2} {\rm g cm}^{-2}$, so any emission lines would have to be formed
close to the surface of the disk and would be suppressed from
optically thin estimates (Storchi-Bergmann~et~al. (1995), Eq. 2).

\section{Long-term Emission}

Because much of the flare's energy is radiated between 10 and 100 eV,
the possibility that the HII region created by the ionizing radiation of
a flare could be observed is intriguing (e.g. Rees 1997).
Assuming that $\sim 5\%$ of the rest mass of the star is radiated in the
flare and using the thick disk spectra discussed above,
the total number of photons between $\sim 100\AA $ and $ 1000\AA$ is
$N_\gamma \sim 10^{63}$.
The ISM surrounding the BH will be in a steady state, because
the recombination time,
$t_{\rm rec} \sim 10^6 n^{-1} T_4^{1/2} {\rm years}$,
is longer than the time between disruptions,
($t_{\rm dis} \sim 10^4 {\rm years}$, where $n$ is the number of atoms
per cm$^{3}$.
The ionizing photon rate is therefore $\sim 3 \times 10^{51} {\rm s}^{-1}$ or
$\sim 1\%$ the total ionizing photon rate from the Galaxy (Mezger 1978;
Bennett~et~al.~1994).
Direct detection of $H_\alpha$ from tidal disruption ionization is therefore
unlikely.

At higher photon energies,
the tidal disruption events may (in a time-averaged sense)
contribute the majority of photons.
The production rate of He~II ionizing photons by tidal disruption
events is $\sim 3 \times 10^{50} {\rm s}^{-1}$,
which if a large fraction (e.g. 10\%)
of the photons ionize He~II and $\sim 20$\% of these ionization
produce He~II~4686, then the luminosity in He~II~4686 (Pacschen-line)
would be $\sim 2 \times 10^{37} {\rm ergs~s}^{-1}$.
The number of He~II photons would be reduced by a factor of $\sim 50$
in the case of the strongest winds described above.
For densities of one atom per cm$^3$,
the radius of the nebula would be
\begin{equation}
R_{\rm HII} \approx \left( {3 N_\gamma \over  4
\pi n^2 \alpha_{\rm B}} \right)^{1/3} \approx 1.5 \times 10^{21}
n^{-2/3} T_4^{1/6} {\rm cm},
\end{equation}
where $\alpha_{\rm B}$ is the case~B recombination coefficient for hydrogen.
A prediction of tidal disruptions is therefore a highly-ionized
nebular emission in the inner kpc of galaxies.
Such nebular emission may be produced by a variety of sources.
He~II~4686 lines for instance may be produced by
Of stars (\cite{ber77}) or Wolf-Rayet stars (e.g. \cite{sch92}),
although such lines would generally be broad ($\sim 3-10\AA$) whereas nebular
lines would be narrow. Possible competing ionizing sources include shocks
(\cite{shu79}), X-ray binaries (\cite{gar90}), and Wolf-Rayet stars.
While all of these objects are concentrated towards galactic centers,
it may be possible to distinguish between a tidal disruption nebula
and most other ionizing sources.
The fossil nebulae may be most readily
observed in galaxies where the background ionizing level is small,
corresponding to a small star formation rate, but
since ample gas is required to absorb the radiation, the best candidates for
detection of the nebulae may be Sa galaxies.
Such nebulae may also be visible in radio lines.

The idea that the unbound debris may produce a supernova remnant has been
discussed by Khokhlov and Melia (1996). For typical disruptions,
the unbound material is emitted in a fan of small opening angle,
$\Omega < 0.1$~rad$^2$, with an energy of $\sim 10^{50}$~ergs.
Rare events with
pericenters close to the Schwarzschild radii, may provide $10^{52}$~ergs.
Such events may produce supernova remnants that could be observed in external
galaxies.
For remnants which sweep up 100 times their initial mass, the
remnants would be displaced from the black hole by
\begin{equation}
R_{\rm disp} \approx 40 \Omega_{0.1}^{-1/3} n^{-1/3} {\rm pc}.
\end{equation}
A supernova remnant lasts $\sim 10^4~{\rm years}$, corresponding to
one such remnant per galaxy.

\section{Conclusions}

Tidal disruption flares occur in two
classes -- those which are sub-Eddington and those which radiate
near the Eddington limit.
Flares from black holes above $\sim 2 \times 10^7 M_\odot$
will generally not radiate above the Eddington limit.

The brightest flares will occur around $\sim 5\times 10^7 M_\odot$
black holes, but they will be relatively rare as they require a finely
tuned impact parameter. The majority of the brightest events will occur
around $\sim 2 \times 10^7 M_\odot$ black holes.
The corresponding timescale for decay of such flares (Eq. \ref{dur}) is about
one month. For all of the spectral scenarios considered, the
optical luminosity is high with peak apparent magnitudes in
U and V of -21 and -19.5 for a $10^7 M_\odot$ black hole
and -19 and -17.5 for a $10^6 M_\odot$ black hole, neglecting reddening.
Both high redshift supernovae searches and low redshift
supernovae searches may observe the flares.
As tidal disruption events occur at a rate of $\sim 100$ times less
frequently than supernovae, many supernovae must be observed before
a flare is expected to be seen. Furthermore, if the flares come from
the centers of galaxies, then they may be harder to detect against the bright
nuclei. The extreme blueness of the flares may help separate them from
the background. In fact, because the spectra are so blue, K-corrections
may actually brighten the flares in optical bands.

The He~II ionizing radiating produced in the flares may dominate that
which is produced in the centers of quiescent galaxies, creating 
a steady state, highly ionized diffuse nebula with extent of $\sim 1$~kpc
which may be observable in recombination lines in all galaxies
with both tidal disruption events and gas near the nucleus.

\acknowledgements

I thank J. Goodman, D. Spergel, and M.J. Rees for helpful discussions.
B. Paczy\'nski's input into this research was invaluable.
AU was supported by an NSF graduate fellowship and NSF
grants AST93-13620 and AST95-30478.

\end{document}